\documentclass[conference]{IEEEtran}
\IEEEoverridecommandlockouts

\usepackage[scr=rsfs]{mathalpha}

\usepackage{graphicx, amsfonts,amsmath , boldline, epstopdf,amssymb }
\usepackage{tabu, multirow,multicol ,booktabs ,setspace ,algcompatible ,mathrsfs,dsfont, array, boldline, makecell, booktabs }
\usepackage{amssymb}
\usepackage[linesnumbered,lined,boxed,commentsnumbered,ruled,longend]{algorithm2e}
\usepackage{adjustbox, stackengine, color, colortbl, epsfig,cite, cases, enumitem, mathtools, tikz, verbatim, bbding}
\usepackage{xcolor}
\definecolor{amethyst}{HTML}{9966CC}
\definecolor{shamrockgreen}{rgb}{0.0, 0.62, 0.38}
\definecolor{selectiveyellow}{rgb}{1.0, 0.73, 0.0}
\newcolumntype{?}{!{\vrule width 1.2pt}} 
\usepackage[caption = false]{subfig}
\usepackage[utf8]{inputenc}
\makeatletter

\begin{document}

\title{\vspace{-.1cm} Micro Energy-Water-Hydrogen Nexus:\\ Data-driven Real-time Optimal Operation\\
\thanks{This work is supported by U.S. National Science Foundation under Award CBET\#2124849.}
}

\author{\IEEEauthorblockN{ Mostafa Goodarzi}
\IEEEauthorblockA{\textit{Electrical and Computer Engineering} \\
\textit{University of Central Florida}\\
Orlando, USA \\
mostafa.goodarzi@ucf.edu}
\and
\IEEEauthorblockN{ Qifeng Li}
\IEEEauthorblockA{\textit{Electrical and Computer Engineering} \\
\textit{University of Central Florida}\\
Orlando, USA \\
Qifeng.Li@ucf.edu}}

\maketitle

\begin{abstract}

This paper extends a new concept of energy-water-hydrogen (EWH) nexus, which was recently developed as a solution for reducing carbon emissions from the generation side of power systems, to the distribution side. Under the concept of distribution-level EWH (micro EWH) nexus, renewable energy sources (RES) are utilized to meet the energy needs of a small community. To avoid the uncertainty caused by RESs, this paper aims to investigate the real-time optimal operation of the micro EWH nexus which is however a challenging optimization problem. First, such a large-scale mixed-integer nonlinear programming problem is relaxed into a mixed-integer convex program (MICP) by leveraging the effective convex-hull relaxation technique. Second, a fast data-driven solution method based on active constraint and integer variable prediction is presented, which can solve the MICP problem very fast since it utilizes historical optimization data to quickly predict binary variable values and a limited set of active constraints. 

\end{abstract}


\allowdisplaybreaks
\section{Introduction}   \label{sec: Intro}
Climate change concerns have led to an increased focus on reducing carbon emissions across various sectors, including transportation, power generation, heating, and industry. Utilization of renewable energy sources (RESs) is a pivotal strategy for decreasing carbon emissions from the power sector, which contributes a quarter of the total emissions in the US  \cite{epa-ghg-website}. However, challenges arise in incorporating RES into the electrical grid due to hosting capacity limitations, potentially causing grid stability issues such as voltage fluctuations and flickers \cite{samet2020deep}. Hence, the complete replacement of all fossil fuel power plants with RESs is presently impractical, necessitating an alternative approach to decrease carbon emissions from these power plants.

A novel concept of energy-water-hydrogen (EWH) nexus was recently developed as a solution for reducing carbon emission from the generation side of power systems \cite{goodarzi2023economic}, where fossil fuel power plants are equipped with carbon capture systems (CCS) and the captured carbon is reused by combining it with hydrogen to create chemical products  \cite{varela2021modeling}. While the required hydrogen can be obtained from diverse sources \cite{mosca2020process}, in the context of the EWH nexus, it comes from green hydrogen which is produced through water electrolysis using RESs and offers an eco-friendly alternative to fossil fuel-based hydrogen production methods \cite{ghandehariun2016life}.

This paper extends the concept of EWH nexus \cite{goodarzi2023economic} to the distribution side for enhancing the sustainability and efficiency of small communities, such as small islands, remote villages, and isolated coastal cities, where the critical infrastructures can be controlled by a single entity. Under the concept of distribution-level EWH nexus, called micro EWH (\textit{m}-EWH) nexus in this paper, RESs are utilized for meeting the energy needs of a small community with a specific emphasis on optimizing energy and water resources to yield more favorable environmental results. Since water electrolysis requires substantial amounts of water, we propose integrating the water network into green hydrogen applications. Besides, since water electrolysis is interconnected with the water distribution system, and water and power systems are linked at the distribution level, our study focuses on the distribution side instead of the generation side, distinguishing it from previous studies.

We should consider uncertainty when discussing green hydrogen applications because RESs are intermittent.
Several studies have assessed RES uncertainty using probabilistic\cite{li2021optimal}, robust\cite{gu2019power}, and stochastic \cite{bayat2023stochastic} optimization models. However, applying these optimization models to large engineering systems is typically challenging due to their extreme scale and computational complexity. This paper proposes to hedge against the uncertainty by solving the optimal operation problem of \textit{m}-EWH in real-time. However, the original problem is computationally challenging since a \textit{m}-EWH nexus generally consists of power and water distribution systems and a green hydrogen system involving a large number of discrete decision variables. First, such a large-scale mixed-integer nonlinear programming problem is relaxed into a mixed-integer convex program (MICP) by leveraging the advanced convex-hull relaxation technique \cite{li2017convex,li2018micro}. To solve the resulting MICP problem in real-time, we introduce a fast data-driven method that is based on active constraints and integer variable prediction (ACIVP) \cite{bertsimas2022online}. The ACIVP-based method maps input parameters into an optimal strategy that includes the optimal values of binary variables and a set of active constraints. Using ACIVP, we can surrogate the original MICP problem with a smaller-scale continuous convex optimization problem that existing solvers can solve rapidly.

The rest of this paper is organized as follows. Section \ref{sec: EWHN} introduces the \textit{m}-EWH nexus concept and its modeling for remote coastal cities/small islands.In Section \ref{sec: SolEWHN}, we detail the approach for real-time operation. Section \ref{sec: Casestudies} encompasses the numerical results and discussion. Finally, Section \ref{sec: Concolusion} provides conclusions and outlines avenues for future work.

\section{Mathematical Modeal of \textit{m}-EWH Nexus}   \label{sec: EWHN}
\begin{figure}[b]
\vspace{-.5cm}
  \centering
{\includegraphics[width=.455\textwidth]{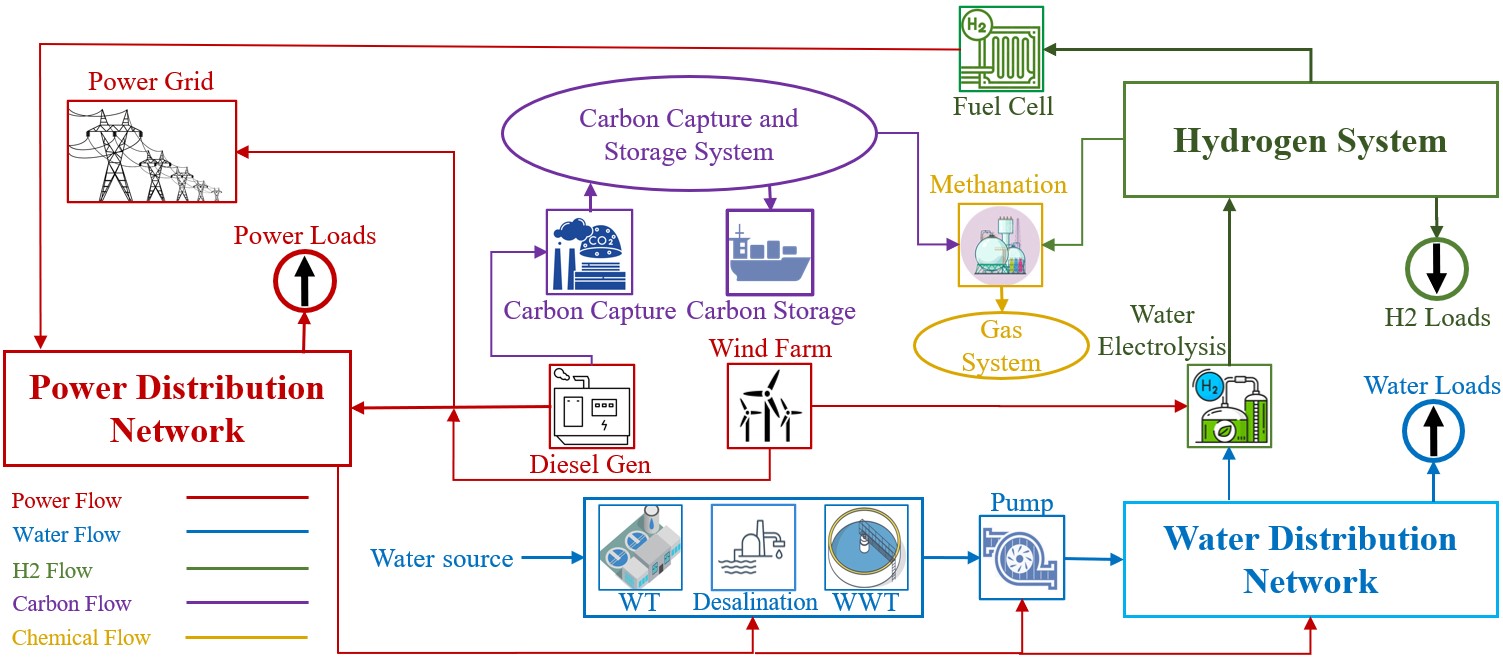}}
 \centering
           \vspace{-.2cm}
\caption{\footnotesize A typical \textit{m}-EWH nexus integrated with a CCS for remote coastal area: \textcolor{red}{\textbf{$\rightarrow$} power}, \textcolor{blue}{\textbf{$\rightarrow$} water}, \textcolor{shamrockgreen}{\textbf{$\rightarrow$} hydrogen}, \textcolor{amethyst}{\textbf{$\rightarrow$} carbon}, \textcolor{selectiveyellow}{\textbf{$\rightarrow$}chemical product}.}
  \label{Fig: EWHN}
\end{figure}

A typical \textit{m}-EWH nexus integrated into a CCS and a methanation system is shown in Fig. \ref{Fig: EWHN}. The \textit{m}-EWH nexus is designed for remote coastal cities/small islands where a single entity controls the entire system. Wind farms provide the energy for electrolyzing water, and excess wind energy is fed into the power distribution system for flexible loads like desalination. Seawater desalination is employed as a promising approach to address water supply challenges in the face of water scarcity \cite{eghtesad2022techno}. In times of surplus wind energy, the power grid can serve as a bulk energy storage system. Hydrogen finds applications in methanation for carbon reuse. It also converts unpredictable wind energy to controllable energy through fuel cell (FC) units and meets hydrogen network demand. The following subsections explain the mathematical models of different components of the proposed system.

\subsection{Power Section} \label{sec:PFPower} 
In this paper, we use the well-known $\mathrm{Distflow}$ model to describe power flow in the distribution system \cite{baran1989optimal}. The convex model of the power section is expressed by \cite{goodarzi2023hybrid,li2018modeling}:\begin{subequations} \label{eq_PDN}
	\begin{align}
 &\sum\limits_{k}(p_{ki,t})+r_{ij}{\mathcal{I}}_{ij,t}-p_{ij,t}=p^\mathrm{dg}_{i,t}+p^\mathrm{sw}_{i,t}+p^\mathrm{hs}_{i,t}-p^\mathrm{l}_{i,t},\label{eq_PDN4}\\
&\sum\limits_{k}(q_{ki,t})+x_{ij}{\mathcal{I}}_{ij,t}-q_{ij,t}=q^\mathrm{dg}_{i,t}+q^\mathrm{sw}_{i,t}+q^\mathrm{hs}_{i,t}-q^\mathrm{l}_{i,t},\label{eq_PDN5}\\
&\mathcal{V}_{i,t}-\mathcal{V}_{j,t}=2(r_{ij}p_{ij,t}+x_{ij}q_{ij,t})-(r^\mathrm{2}_{ij}+x^\mathrm{2}_{ij}) {\mathcal{I}}_{ij,t}, \label{eq_PDN1}\\
&p_{ij,t}^2+q_{ij,t}^2 \leq \mathcal{V}_{i,t}{\mathcal{I}}_{ij,t},\label{eq_PDN2}\\
&\underline{\mathcal{V}}_{i} \overline{\mathcal{V}}_{i} {\mathcal{I}}_{ij,t} + {\overline{s}_{ij}}^2 \mathcal{V}_{i,t} \leq {\overline{s}_{ij}}^2 (\underline{\mathcal{V}}_{i} + \overline{\mathcal{V}}_{i})\label{eq_PDN2_2}\\
&p_{ij,t}^2+q_{ij,t}^2\leq {\overline{s}_{ij}}^2,\label{eq_PDN3}\\
&\resizebox{.89\hsize}{!}{$\underline{p}^\mathrm{dg}_{i},\underline{q}^\mathrm{dg}_{i},\underline{\mathcal{V}}_{i},\underline{\mathcal{I}}_{ij} \leq p^\mathrm{dg}_{i,t},q^\mathrm{dg}_{i,t},\mathcal{V}_{i,t},\mathcal{I}_{ij,t}\leq \overline{p}^\mathrm{dg}_{i},\overline{q}^\mathrm{dg}_{i},\overline{\mathcal{V}}_{i},\overline{\mathcal{I}}_{ij}$}\label{eq_PDN6}\\
    &p^\textrm{tr}_t = p^\textrm{wind}_t - (p^\textrm{we}_t + p^\textrm{sw}_t), \label{eq_netPower} 
	\end{align}
\end{subequations}
where $k$, $i$, and $j$ denote bus numbers, and $t$ represents time. $\overline{s}$, $\mathcal{I}$, $p$ and $q$, and $r$ and $x$ represent the maximum apparent power, square current, active and reactive power flow, and resistance and reactance of the branches, respectively. $\mathcal{V}$, $p^\mathrm{dg}$, $q^\mathrm{dg}$, $p^\mathrm{sw}$, $q^\mathrm{sw}$, $p^\mathrm{hs}$, $q^\mathrm{hs}$, $p^\mathrm{l}$, and $q^\mathrm{l}$ are the square voltage, and active and reactive power of diesel generation, surplus wind, hydrogen system, and power load at each bus, respectively. The power grid can capture excess wind power, as shown with (\ref{eq_netPower}). In this constraint,  $p^\textrm{tr}$, $p^\textrm{wind}$ and  $p^\textrm{we}$ represent the transferred power to the power network, wind power, and water electrolysis power, respectively.

\subsection{Water Section} \label{sec:PFWater}
A convex-hull model for the water section of the \textit{m}-EWH nexus, including water desalination, mass flow conservation law, pipe flow, water pumps, water tank, and pressure-reducing valves, is given \cite{goodarzi2022evaluate}: \begin{subequations} \label{eq3}
	\begin{align}
   &\sum\limits_{m}f_{nm,t} = f^\mathrm{des}_{n,t}-d_{n,t}+f^\mathrm{wt}_{n,t}, \label{eq3_1}\\
   &\widehat{Y}
\begin{cases}
      \leq (2\sqrt{2}-2)r^\mathrm{w}_{q}\overline{f}_{q}f_{q,t}+(3-2\sqrt{2})r^\mathrm{w}_{q}\overline{f}_{q}^2, \\
      \geq (2\sqrt{2}-2)r^\mathrm{w}_{q}\underline{f}_{q}f_{q,t}-(3-2\sqrt{2})r^\mathrm{w}_{q}\underline{f}_{q}^2,  \\
      \geq 2r^\mathrm{w}_{q}\overline{f}_{q}f_{q,t}-r^\mathrm{w}_{q}\overline{f}_{q}^2,  \\
      \leq 2r^\mathrm{w}_{q}\underline{f}_{q}f_{q,t}+r^\mathrm{w}_{q}\underline{f}_{q}^2,
    \end{cases} \label{eq3_2}\\
&\resizebox{.885\hsize}{!}{$p^\mathrm{des}_t = b^\mathrm{des}_t e^\mathrm{des}_\mu  f^\textrm{des}_t, \,\, 0.25(\mu-1) \overline{f}^\textrm{des} \leq f^\textrm{des}_t \leq 0.25\mu \overline{f}^\textrm{des}$},\label{eq_Water_1}\\
&\widehat{Y}+y^{\mathrm{G}}_{q,t}-r^\mathrm{w}_{q}(f_{q,t})^2 \ge M(b^\mathrm{p}_{q,t}-1),\label{Convwatp1}\\
&\widehat{Y}+y^{\mathrm{G}}_{q,t}-r^\mathrm{w}_{q}\overline{f}_{q}f_{q,t} \le M(1-b^\mathrm{p}_{q,t}),\label{Convwatp2}\\
&0 \le f_{q} \le b^\mathrm{p}_{q,t} \overline{f}_{q}.\label{Convwatp3}\\
&V_{n,t+1}^\mathrm{wt}=V_{n,t}^\mathrm{tw}+f_{n,t}^\mathrm{wt}, \label{eq3_6}\\      	
&A_{n}^\mathrm{wt} (y_{n,t+1}^\mathrm{wt} - y_{n,t}^\mathrm{wt}) = f_{n,t}^\mathrm{wt}, \label{eq3_7}\\
&-PR \leq \widehat{Y} \leq PR, \label{eq3_8}\\
&\eta p^\mathrm{p}_{i,t} \ge 2.725\times (a_1(f_{q,t})^2+a_0f_{q,t})\label{ConvPump_1}\\
&\eta p^\mathrm{p}_{i,t} \le 2.725\times (a_1\overline f_{q}+a_0) f_{q,t},\label{ConvPump_2}\\
&\underline{f}^\mathrm{des}_{n}, \underline{f}^\mathrm{wt}_{n},\underline{V}^\mathrm{wt}_{n} \leq f^\mathrm{des}_{n,t}, f^\mathrm{wt}_{n,t},V^\mathrm{wt}_{n,t} \leq \overline{f}^\mathrm{des}_{n}, \overline{f}^\mathrm{wt}_{n},\overline{V}^\mathrm{wt}_{n},\label{eq3_4}\\
&\underline{y}_{n}, \underline{f}_{p} \leq y_{n,t}, f_{p,t},\leq \overline{y}_{n}, \overline{f}_{p},\label{eq3_5}
	\end{align}
\end{subequations}
where $\widehat{Y}= y^c_{n,t}-y^c_{m,t}+h_{q}$, $\mu \in \{1,2,3,4\}$, and $q$ is the pipe between node $n$ and node $m$. The equality of water injection and water output at each node is guaranteed by (\ref{eq3_1}), where $f^\mathrm{des}$, $f^\mathrm{wt}$, and $d$ are the water flow of desalination, water flow of the tank, and water demand, respectively. Constraint (\ref{eq3_2}) shows a convex-hull model for head loss along a regular pipe where $r^\mathrm{w}$ is the head loss coefficient of the pipe. Equation (\ref{eq_Water_1}) shows the power demand for water desalination. The convex model of a pipe with a pump is expressed by (\ref{Convwatp1}) to (\ref{Convwatp3}), where $y^{\mathrm{G}}$ and $b^\mathrm{p}$ represent head gain and the binary variable related to the pump. Each tank is modeled as a node using (\ref{eq3_6}) and (\ref{eq3_7}), where $V^\mathrm{wt}$ is the water tank volume. A pressure-reducing valve is modeled by (\ref{eq3_8}) to control water head pressure, and the convex model of a pump is modeled by (\ref{ConvPump_1}) and (\ref{ConvPump_2}).

\subsection{Hydrogen Section}\label{sec:Hydrogen}
In the proposed \textit{m}-EWH nexus, proton exchange membrane electrolysis is selected due to its flexibility. The following mathematical formulations describe the hydrogen section of the \textit{m}-EWH nexus: \begin{subequations} \label{eq_H2}
	\begin{align}
&h^\textrm{we}_t = \xi^\textrm{we}_\textrm{p} p^\textrm{we}_t \label{eq_H2_1}\\
&d^\textrm{we}_t = \xi^\textrm{we}_\textrm{w}  h^\textrm{we}_t,\label{eq_H2_2}\\
&p^\textrm{fc}_t = \xi^\textrm{fc}_\textrm{h} h^\textrm{fc}_t,\label{eq_H2_4}\\
&(p^\textrm{we}_t - p^\textrm{fc}_t)^2 + (q^\textrm{hs}_t)^2 \leq (\overline{s}^\textrm{hs})^2,\label{eq_H2_5}\\
&b^\textrm{fc}_t\underline{h}^\textrm{fc}, b^\textrm{we}_t\underline{p}^\textrm{we} \le h^\textrm{fc}_t, {p}^\textrm{we}_t \leq b^\textrm{fc}_t\overline{h}^\textrm{fc}, b^\textrm{we}_t\overline{p}^\textrm{we},\label{eq_H2_6}\\
&b^\textrm{we}_t + b^\textrm{fc}_t \leq 1,\label{eq_H2_8}
    \end{align}
\end{subequations}
where $h^\textrm{we}$, $h^\textrm{fc}$, $p^\textrm{we}$ and $p^\textrm{fc}$ show the amount of hydrogen and power of the electrolysis and FC system. $\xi$ is constant and $b^\textrm{we}$ and $b^\textrm{fc}$ represent the binary variables regarding electrolysis and FC. Constraint (\ref{eq_H2_1}) shows hydrogen production and (\ref{eq_H2_2}) represents the required water for the electrolysis process. Constraint (\ref{eq_H2_4}) refers to the hydrogen consumption level of FC units based on conversion factors, and (\ref{eq_H2_5}) presents the hydrogen systems' inverter for reactive power support \cite{haggi2022proactive}. The simultaneous operation of electrolysis and FC units is avoided by (\ref{eq_H2_8}).

\subsection{CCS and Methanation Section}\label{sec:CSSM}
The diesel generator of the proposed system is required to employ a CCS for carbon emission reduction. Following carbon capture, the methanation system integrates the captured carbon with hydrogen to produce $CH_4$ through a Sabatier reaction \cite{li2018optimal}. Following is a model of this section:
\begin{subequations} \label{eq_CCSM}
	\begin{align}
&c_t^\textrm{dg} = \xi^\textrm{dg}_\textrm{c}p^\textrm{dg}_t,\label{eq_CCS_1}\\
&c^\textrm{e}_t = c_t^\textrm{dg} - c_t^\textrm{s} - c^{\chi}_t,\label{eq_CCS_2}\\
&I^\textrm{$\chi$}_t = \rho^\textrm{$\chi$} \, \xi_\textrm{c}^\textrm{$\chi$} \, c^\textrm{$\chi$}_t, \label{eq_Chemical_S}
	\end{align}
\end{subequations}
where $c^\textrm{dg}$ represents the carbon emissions of the diesel generator, consisting of three parts: $c^\textrm{e}$ for emitted portions, $c^{\chi}$ for those reused for chemical production, and $c^\textrm{s}$ for those stored \cite{akbari2021economic}. $I^\textrm{$\chi$}$ is the revenue from selling the chemical product, and $\rho^\textrm{$\chi$}$ is the unit price of the chemical product.

\section{Solution Method for Real-time Operation}   \label{sec: SolEWHN}
\begin{figure}[t]
  \centering
{\includegraphics[width=.489\textwidth]{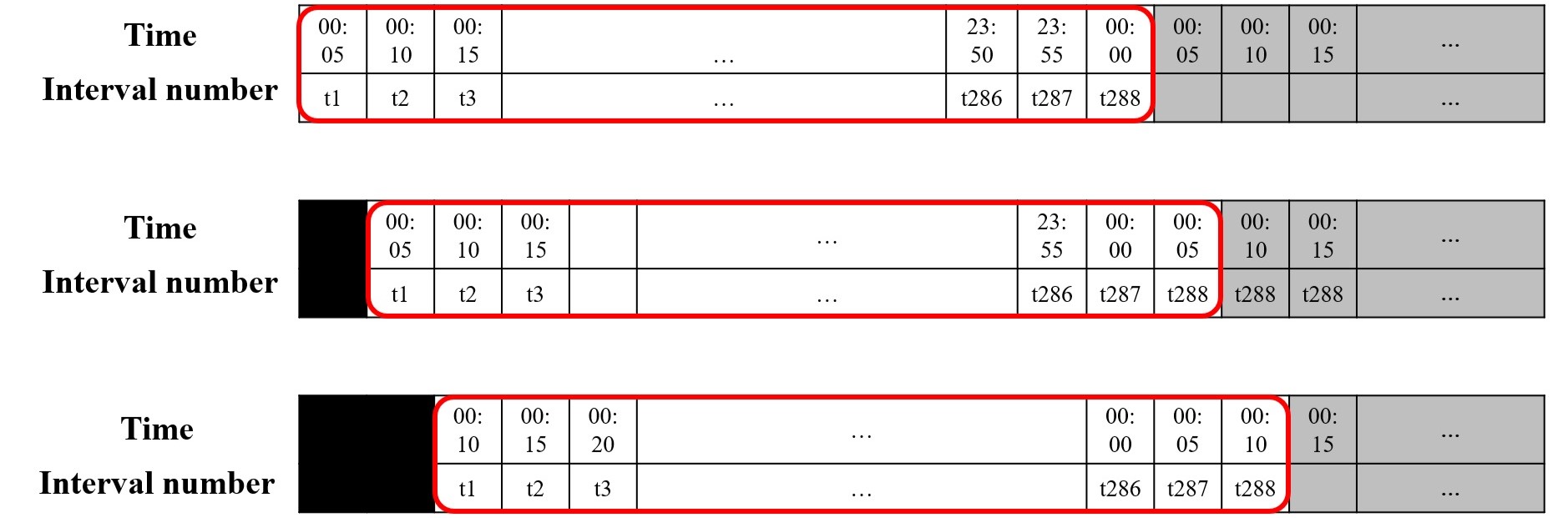}}
 \centering
           \vspace{-.4cm}
\caption{\footnotesize Real-time optimization model for the \textit{m}-EWH nexus.}
          \vspace{-.7cm}
  \label{Fig: roll}
\end{figure}

\begin{figure}[b]
\vspace{-.6cm}
  \centering
{\includegraphics[width=.489\textwidth]{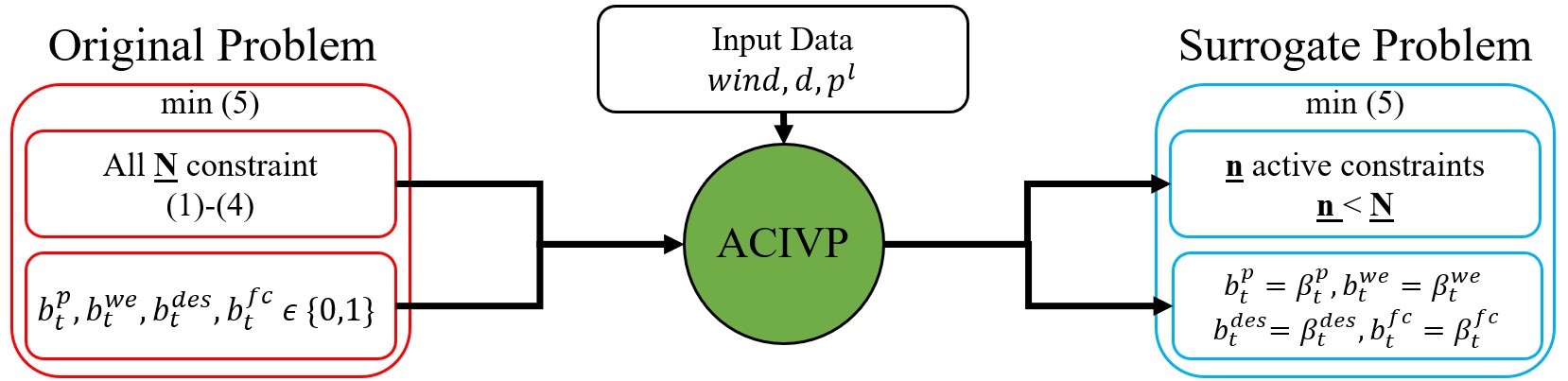}}
 \centering
           \vspace{-.4cm}
\caption{Surrogate the original problem using ACIVP-based method: input data are wind, water demand, and power demand}
  \label{Fig: surr}
\end{figure}

The objective function of these optimization models that minimize carbon emissions and wind power transfer to the power grid, while simultaneously maximizing revenue from the sale of chemical production is shown as:
\begin{align} \label{eq_Objective} 
    \sum_{t=1}^{T}\,\,\biggl( \alpha_1 p^\mathrm{tr}_t + \alpha_2 p^\mathrm{dg}_{t} + \alpha_3 c_t^\textrm{e} + \alpha_4 c_t^\textrm{s} 
 -  I^\textrm{$\chi$}_t \biggl),
\end{align}
where $\alpha_1$ to $\alpha_4$ are the parameters that show the penalty \cite{shao2019low} or cost. 
The optimization model for the \textit{m}-EWH nexus should consider operational horizons of 24 hours since water and hydrogen tanks can store and utilize these resources throughout the day. Additionally, the proposed system depends on actual wind speed to manage wind uncertainty in real-time optimal operation mode. This study assumes that wind predictions for the next 5 minutes are accurate enough to be treated as real-time wind speed. We propose using a rolling window and solving an optimization problem every 5 minutes for 24 hours, as shown in Fig. \ref{Fig: roll}. Following is the optimization problem for the \textit{m}-EWH nexus:
\begin{align} \label{eq: OpMo}
& \textbf{min} \,\,\,\,\ (\ref{eq_Objective}) \nonumber  \\
& s.t \,\,\,\,\,\,\,\ (\ref{eq_PDN}) - (\ref{eq_CCSM}) \nonumber  \\
& \,\,\,\,\,\,\,\,\,\,\,\,\,\,\,\, b_t^\textrm{p}, b_t^\textrm{we},b_t^\textrm{des},b_t^\textrm{fc} \in \{0,1\}.
	\end{align}
Due to the requirement to solve (\ref{eq: OpMo}) every five minutes, the computational efficiency of the solution method is critical. Even though we have convexified the non-convex constraints in these problems, the binary variables and the large size still make the problem computationally challenging. We introduce the ACIVP-based approach \cite{bertsimas2022online}, which is a novel and fast solution method for solving MICP problems, to find the real-time optimal operation of the \textit{m}-EWH nexus. By employing data-driven methods, ACIVP predicts the active constraints and optimal values of binary variables and replaces (\ref{eq: OpMo}) with a surrogate optimization problem containing only active constraints and optimal values of binary variables. Data from solved offline optimization problems are used for this purpose. Fig. \ref{Fig: surr} shows the ACIVP-based method where $\beta$s are the optimal values of the binary variables.
The surrogate optimization problem is much easier than the original one because it has a smaller number of constraints and contains only continuous, allowing us to apply MICP to real-time operations that were previously impractical. ACIVP uses a machine learning technique that performs well at classification to map the input data, including real-time and forecasted wind speeds, water demand, and power demand values, to an optimal strategy including $\beta$s and active constraints: 
\begin{equation} \label{eq:superMLF}
\psi = \mathcal{H}(\varphi),
\end{equation}
where $\varphi$ and $\psi$ are the input and output of the prediction model, respectively. The prediction model uses $\Gamma =\{(\varphi_1,\psi_1),(\varphi_2,\psi_2),...,(\varphi_{\mathcal{N}_\mathrm{d}},\psi_{\mathcal{N}_\mathrm{d}})\}$ as a dataset to do the training task. The training dataset, which is created by solving offline optimization problems, is required to develop the hypothesis function. Algorithm \ref{alg:OCTsOMIO} shows the proposed solution process of the real-time optimal operation of the \textit{m}-EWH nexus that uses the ACIVP-based fast solution method. The input data includes the set of the wind speed, water demand, and power demand and $\psi= \{\beta_t^\textrm{p}, \beta_t^\textrm{we},\beta_t^\textrm{des},\beta_t^\textrm{fc},\gamma_t^\textrm{S}\}$ represents the related strategy.
\begin{algorithm}[!h]
\label{alg:OCTsOMIO}
  \caption{\footnotesize ACIVP-based method for \textit{m}-EWH nexus}
  \footnotesize 
\begin{algorithmic}[1]
\STATE Input data: wind speed, water demand, and power loads ($\varphi$)
\STATE Use (\ref{eq:superMLF}) to find the optimal values for binary variables and active constraints $\psi= \{\beta_t^\textrm{p}, \beta_t^\textrm{we},\beta_t^\textrm{des},\beta_t^\textrm{fc},\gamma_t^\textrm{S}\}$.
\STATE Set the integer variables to the optimal values and update (\ref{eq_Water_1}) - (\ref{Convwatp3}), and (\ref{eq_H2_6}). Remove the redundant constraints considering $\gamma_t^\textrm{S}$.
\STATE Develop the surrogate optimization problem.
\STATE Solve the surrogate optimization problem, which is smaller and continues, to obtain the optimal operation of the \textit{m}-EWH nexus.
\end{algorithmic}
\end{algorithm}

\section{Numerical Results and Discussion} \label{sec: Casestudies}
\begin{figure}[!b]  
    \vspace{-.6cm}
  \centering
\includegraphics[width=0.35\textwidth]{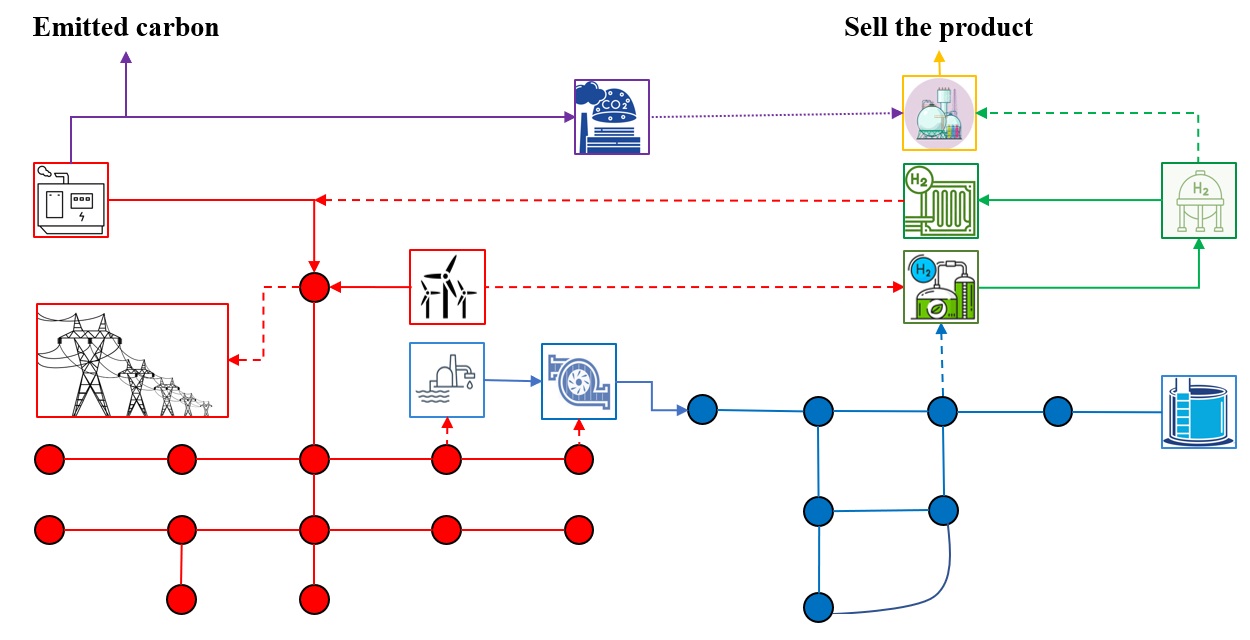}
 \centering
     \vspace{-.2cm}
    \caption{The \textit{m}-EWH nexus for a small stand-alone coastal city.}
  \label{pic: EWHNIEEE13} 
\end{figure}

\begin{figure}[t]
  \centering
  \subfloat[]{\includegraphics[width=0.35\textwidth]{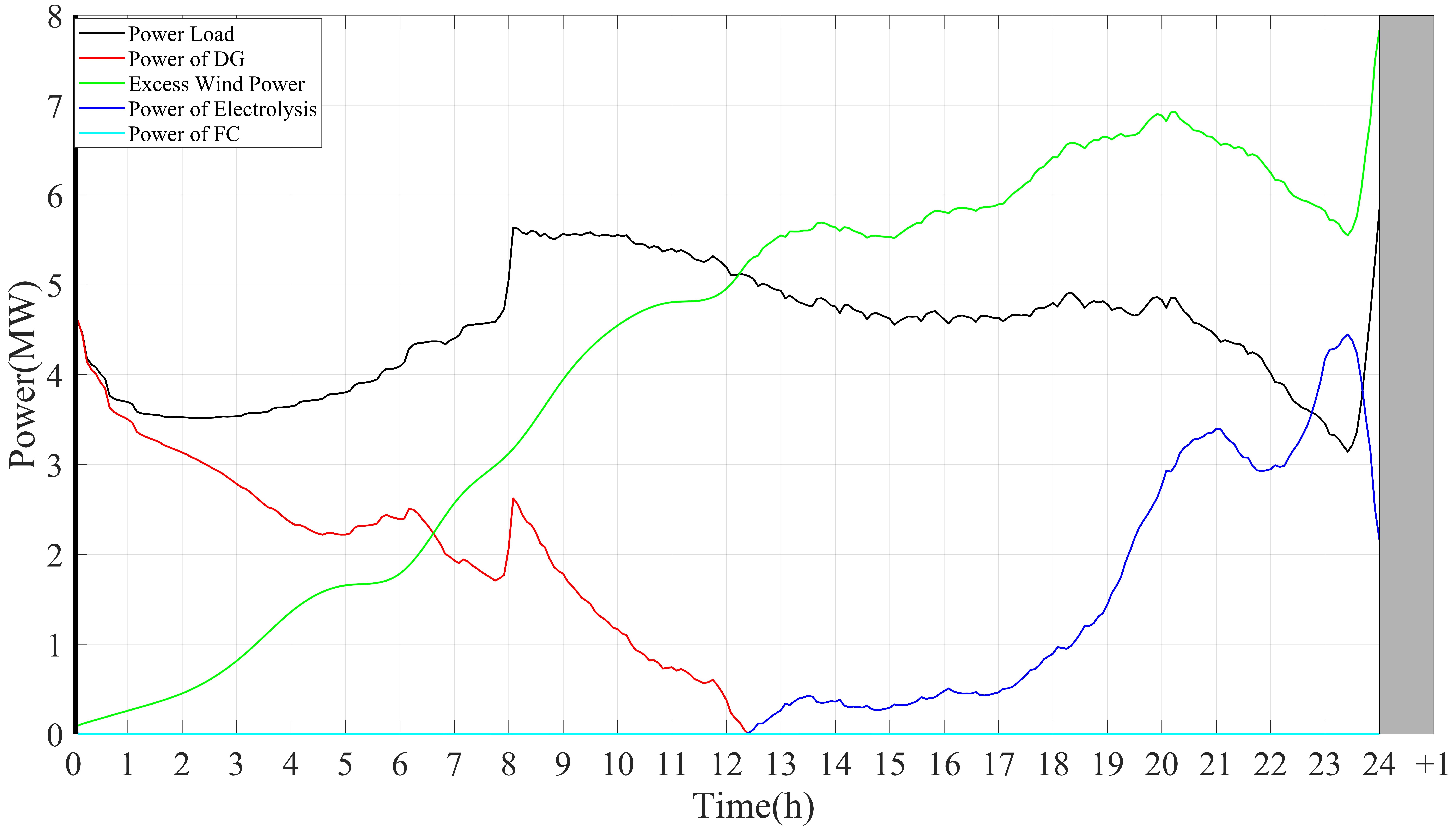} \label{pic: casestudy1}}\\  
  \vspace{-.4cm}
  \subfloat[]{\includegraphics[width=0.35\textwidth]{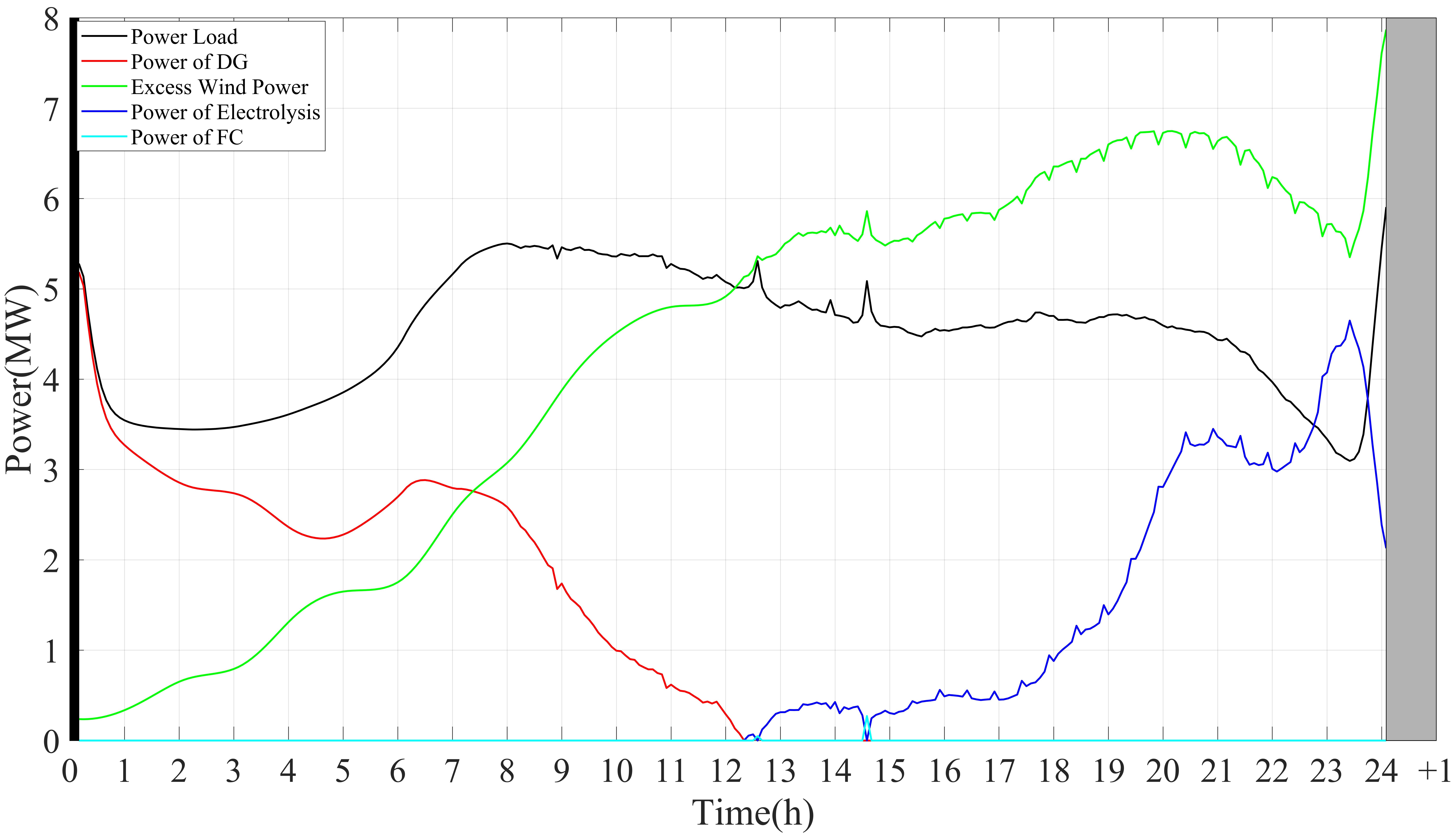}\label{pic: casestudy2}}\\
  \vspace{-.4cm}
  \subfloat[]{\includegraphics[width=0.35\textwidth]{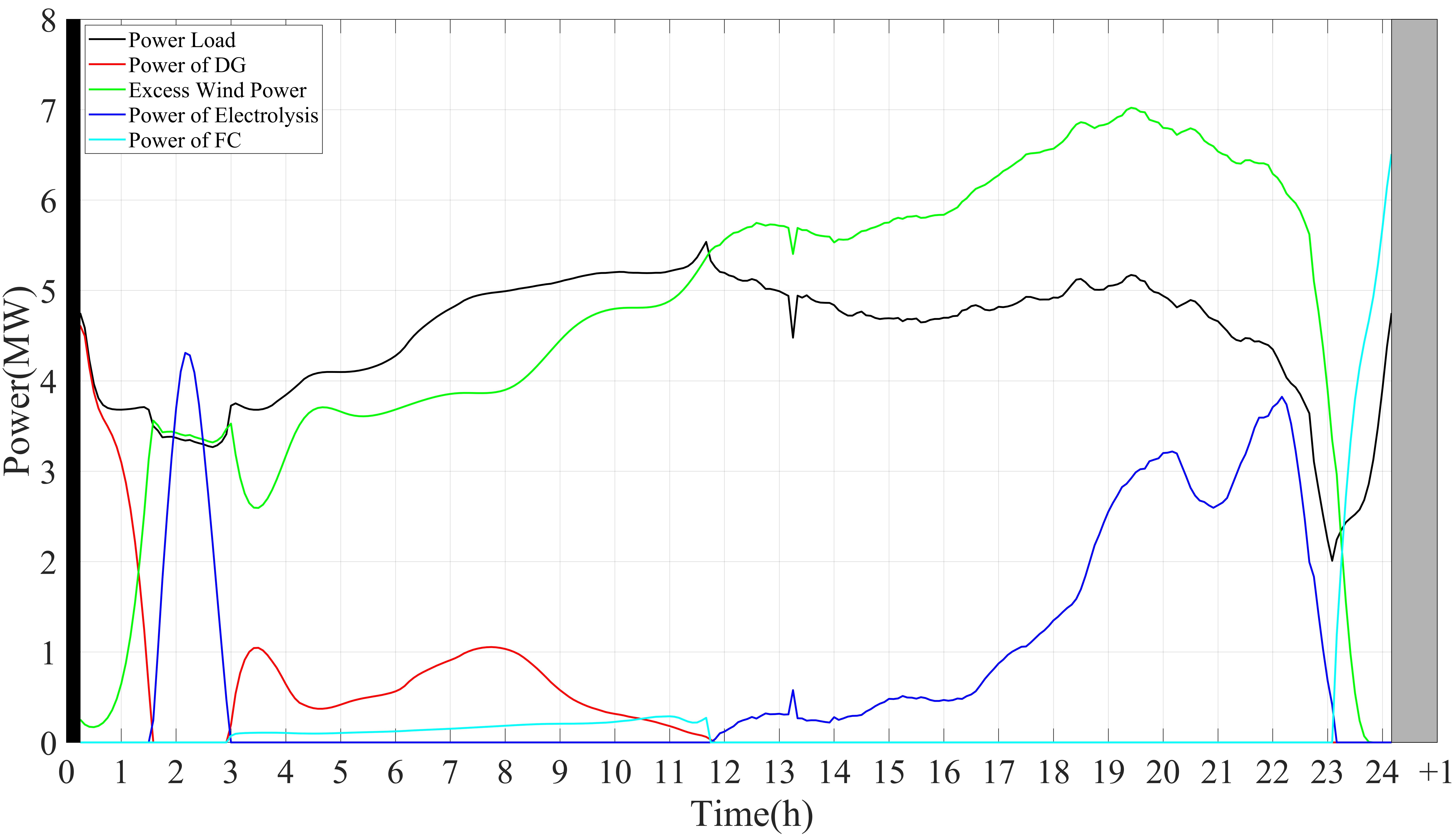}\label{pic: casestudy3}}\\
  \vspace{-.2cm}
  \caption{\textit{m}-EWH nexus optimal operation in three consecutive time steps: \\ a) 00:05 b) 00:10, c) 00:15.} 
  \label{pic: casestudy}
      \vspace{-.5cm}
\end{figure}

We assess the resilience of our proposed method using a modified IEEE 13-bus system integrated with an 8-node EPANET water system \cite{goodarzi2022security}, representing a small standalone coastal city (Fig. \ref{pic: EWHNIEEE13}). 
\begin{figure}[!b]  
    \vspace{-.2cm}
  \centering
\includegraphics[width=0.35\textwidth]{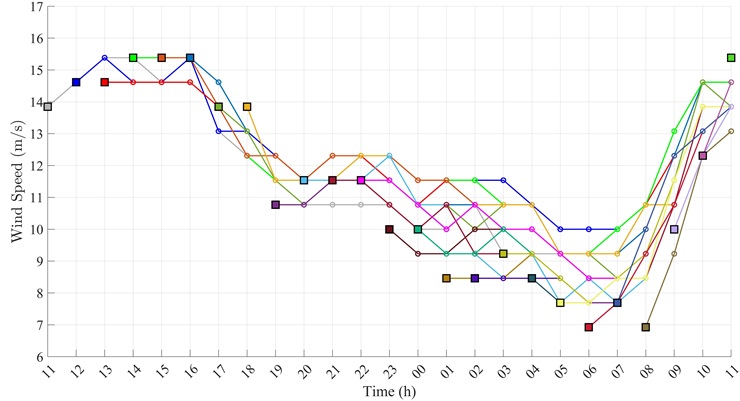}
 \centering
     \vspace{-.2cm}
    \caption{Actual wind speed $\blacksquare$ and forecasted wind speed -o-.}  \label{pic: windforc}    
\end{figure} 
Training data sets are generated from offline optimization problems based on real wind speed and load data. The 24-hour wind speed curves from 2008 to 2022 \cite{meteoblue} and the nodal load curves (excluding water pumps and desalination) \cite{dataminer2} are extracted and interpolated at five-minute intervals. The ACIVP-based method is evaluated using real data, including actual wind speed and load data from the Midcontinent Independent System Operator (MISO) \cite{Misowind}. We employ this data to optimize the \textit{m}-EWH nexus operation, comparing solution times between the conventional method and the ACIVP-based method. According to Fig. \ref{pic: windforc}, even the one-hour-ahead prediction of wind speed is inaccurate, indicating the need for the optimization problem to be solved rapidly. We utilize the real-time optimization models discussed in the preceding section to analyze the case study (depicted in Fig. \ref{pic: EWHNIEEE13}) and derive the optimal operation for the \textit{m}-EWH nexus. The results for three consecutive time intervals are illustrated in Fig. \ref{pic: casestudy}. 
This figure emphasizes the importance of real-time optimal operation for the \textit{m}-EWH nexus. The observed differences between forecasted and actual wind speeds significantly affect the optimal operation of the \textit{m}-EWH nexus. In this figure, the power loads are depicted in black, including the water pump, water desalination, and other power loads connected to the power distribution network. Power generation by diesel generators is shown in red.
The green color represents the surplus wind energy after meeting the requirements for water electrolysis. The power demand for water electrolysis is indicated by the blue color, while the power of the FC is represented by the cyan color. Fig. \ref{pic: casestudy3} indicates that by altering the wind speed prediction value at 00:15 for the time between 01:00 to 03:00, water electrolysis should be turned on from 01:30 to 03:00. In the two previous intervals (00:05 and 00:10), water electrolysis had been off according to the optimal operation (Fig. \ref{pic: casestudy1} and Fig. \ref{pic: casestudy2}). Therefore, modifying the wind speed prediction during the third time interval (00:15) changes the optimal operation of water electrolysis, consequently impacting the overall operation of the \textit{m}-EWH nexus. It is critical to quickly resolve the optimization problem to obtain real-time optimal operation of the \textit{m}-EWH nexus because any discrepancy between predicted and actual wind speeds significantly impacts its optimal operation. Solving the related optimization problems using conventional approaches takes over 9 to 16 minutes. The proposed ACIVP-based method achieves optimal real-time operation of the \textit{m}-EWH nexus in just a few seconds by replacing the initial problem with a surrogate one. Table \ref{tab: sol} provides a comparison of solution times between conventional approaches and the ACIVP-based method. This table demonstrates that the ACIVP-based method enables us to effectively manage wind energy uncertainty and solve the real-time optimization problem of the \textit{m}-EWH nexus.


\begin{table}[!t]
\centering
\footnotesize
\captionsetup{labelsep=space,font={footnotesize,sc}}
\caption{\textrm{comparison of solution times between conventional methods and the ACIVP-based method}}
\vspace{-.2cm}
\footnotesize
\begin{tabular}{ccc}
\hline \hline
\multirow{2}{*}{Time Step} &  \multicolumn{2}{c}{Solution Time (sec)}   \\ 
&  Conventional Method  & ACIVP-based Method   \\ 
\midrule
00:05 & 592.70   & 1.24 \\
00:10 & 562.99   & 1.31 \\
00:15 & 984.50   & 1.15 \\
\midrule
\midrule
\end{tabular}
\label{tab: sol}
\vspace{-.5cm}
\end{table}

\section{Conclusion and Future Work} \label{sec: Concolusion}
This paper introduces real-time optimization models for the \textit{m}-EWH nexus, encompassing power, water, and hydrogen systems, to reduce carbon emissions while maximizing RESs utilization. The \textit{m}-EWH nexus leverages wind power for hydrogen generation through electrolysis, combining it with captured carbon to mitigate carbon emissions from the power sector. The paper formulates a convex mathematical model for the optimal operation of the \textit{m}-EWH nexus, which poses computational challenges. The ACIVP-based method based on historical optimization data is introduced to address this issue. It predicts binary variable values and active constraints based on historical optimization data. The ACIVP-based method replaces the initial problem with a small-scale continuous convex optimization problem to achieve rapid resolution. Validation is conducted through a case study of the \textit{m}-EWH nexus, showcasing a significant reduction in solution time, enabling real-time optimal operation in the face of intermittent wind energy and power demand.

Future work could involve exploring other electrolysis technologies, such as Alkaline water electrolysis, which presents a more challenging problem. Additionally, the \textit{m}-EWH nexus, as a novel engineering problem, holds potential for expansion into other sectors, particularly the transportation system, a major contributor to total emissions in the US, where green hydrogen can reduce carbon emissions. ACIVP-based methods can also be enhanced to increase their efficiency by increasing the accuracy of the data-driven part.

\bibliographystyle{IEEEtran}	
\bibliography{Main}

\end{document}